\documentstyle [prl,aps,multicol,amstex,epsfig]{revtex}
\begin{document}
\hyphenation{following}
\title{ Group-Theoretical Analysis of Second Harmonic Generation at (110) and 
(111) Surfaces of Antiferromagnets }
\author{M. Trzeciecki$^{1,2}$ and W. H\"ubner$^1$}
\address{$^1$Max-Planck-Institut f\"ur Mikrostrukturphysik, Weinberg 2,
D-06120 Halle, Germany}
\address{$^2$Institute of Physics, Warsaw University of Technology, Koszykowa 
75, 00-662 Warsaw, Poland}

\date{\today}
\maketitle

\begin{abstract}
Extending our previous work we classify the nonlinear magneto-optical response 
at low index surfaces of fcc antiferromagnets, such as NiO. Antiferromagnetic 
bilayers are discussed here as models for the termination of bulk 
antiferromagnets.
\end{abstract}
\pacs{78.20.Ls, 75.30.Pd, 75.50.Ee, 42.65.-k}

\begin{multicols} {2}
\section{Introduction}

Nonlinear optics has been proven to be very useful for the investigation of 
ferromagnetism at surfaces due to its enhanced sensitivity to 
twodimensional {\em ferromagnetism} \cite {ref53}. The magnetic effects are 
usually much stronger than in linear optics (rotations up to 90$^\circ$, 
pronounced spin polarized quantum well state oscillations 
\cite{ref64,ref65,vol1}, 
magnetic contrasts close to 100$\%$) \cite {vol1,ref22,pustu,vol2}. Recently, 
Second Harmonic Generation (SHG) has been successfully applied to probe {\em 
antiferromagnetism} (visualization of bulk AF domains \cite {ref18,ref4,ref12}). 
The potential of SHG to study the surface antiferromagnetism has been announced 
in Ref. \cite{dahn} extensively discussed in our previous paper \cite{ourpaper}.

The practical importance of studies in this field follows from the applications 
of antiferromagnetic (AF) oxide layers in devices such as those based on TMR 
(tunneling magnetoresistance), where a trilayer structure is commonly 
used. The central layer 
of TMR devices consists of an oxide sandwiched between a soft and a hard 
magnetic layer. For these technological applications it is necessary 
to develop a technique to study buried oxide interfaces. Such a technique can be 
SHG. One of the promising materials for the mentioned devices is NiO. 
However, to the best of our knowledge, the understanding of its detailed spin 
structure is scarce - even the spin orientation on the ferromagnetically ordered 
(111) surfaces is not known.

Our recent paper \cite{ourpaper} presented an extensive study of the nonlinear 
electrical susceptibility tensor $\chi^{(2\omega)}_{el}$ (the source for SHG 
within the electrical dipole approximation), mostly for monolayer structures. It 
has been proven there that the spin structure of an antiferromagnetic monolayer 
can be detected by means of SHG. Also the possibility of antiferromagnetic 
surface domain imaging has been presented for the first time. As it was 
mentioned in this previous work, bilayer spin structures are enough to account 
for the symmetry of a surface of a cubic antiferromagnet. Here we present an 
extention of that work to the (110) and (111) bilayer structures, thus 
completing our group theoretical analysis of low Miller-index antiferromagnetic 
surfaces. 

\section{Results}
We follow exactly the group theoretical method described in Ref. 
\cite{ourpaper}. At this point it is necessary to define the notions of 
``phase'' and ``configuration'', used henceforth to classify our results. 
``Phase'' describes the magnetic phase of the material, i.e. paramagnetic, 
ferromagnetic, or AF. Secondly, the word ``configuration'' is reserved for the 
description of the magnetic ordering 
of the surface. It describes the various possibilities of the spin ordering, 
which are different in the sense of topology. We describe AF 
configurations, denoted by little letters a) to l), as well as several 
ferromagnetic configurations, denoted as ``ferro1'', ``ferro2'', etc. The number 
of possible configurations varies depending on the surface orientation. All the 
analysis concerns collinear antiferromagnets, with one easy axis.

The tables show the allowed tensor elements for each configuration. The 
tables also contain the information on the parity of the nonvanishing tensor 
elements: the odd ones are printed in 
boldface. In some situations an even tensor element (shown in lightface) is 
equal to an odd element (shown in boldface), this means that this pair of tensor 
elements is equal in the domain which is depicted on the corresponding figure, 
but they are of opposite sign in the other domain. The parity of the elements 
has been checked in the operations $2_z$, $4_z$, and in the operation connecting 
mirror-domains to each other (for the definition of the mirror-domain structure 
see Ref. \cite{ourpaper}). The domain operation(s) on which the parity depends 
is (are), 
if applicable, also displayed in the tables. If two or more domain operations 
have the same effect, we display all of them together. To make the tables 
shorter and more easily readable some domain operations (and the corresponding 
parity information for the tensor elements) are not displayed, namely those that 
can be created by a superposition of the displayed domain operations. We also do 
not address the parity of tensor elements in the $6_z$ nor $3_z$ operations for 
(111) surfaces nor any other operation that ``splits'' tensor elements, although 
these operations also lead to a domain structure \cite{par111}. As has been 
discussed in Ref. \cite{ourpaper} it is possible to define a parity of the 
tensor elements for the $3_z$ and $6_z$ operations, however the tensor elements 
then undergo more complicated changes. The situations where the parity of the 
tensor elements is too complicated to be displayed in the tables are indicated 
by a hyphen in the column ``domain operation''. For the paramagnetic phase, 
where no domains exist, we display the hyphen as well. For some configurations, 
none of the operations leads to a domain structure - in those configurations we 
display the information ``one domain''. The reader is referred to Ref. 
\cite{ourpaper} for the particularities of the parity check.

\subsection{(110) bilayer}
The previously described AF configurations of the (001) monolayer most commonly 
get split into two different configurations when a bilayer structure is 
considered. For the (110) bilayer it is not the case - only two of twelve AF 
configurations get split in this way, thus one obtains 14 AF configurations of 
the (110) bilayer. Describing the results of our analysis we use the 
nomenclature of our previous article, i.e. the antiferromagnetic configurations 
are labeled by small letters. Only the four configurations that result from 
splitting of the two configurations of the monolayer structure are labeled by 
small letters with subscripts that carry the information about how they have 
been constructed from the (110) monolayer. For configurations with subscript 
``a'' the lower layer is constructed by translation of the topmost layer by 
vector (0.5a, 0.5b), where a and b are interatomic distances {\em within} the 
(110) plane along $x$ and $y$ axes, respectively. For configurations with 
subsript ``b'' the vector of translation is (-0.5a, 0.5b). This corresponds to 
the way we constructed the (001) bilayers in \cite{ourpaper}. 

\begin{figure}
\epsfig{file=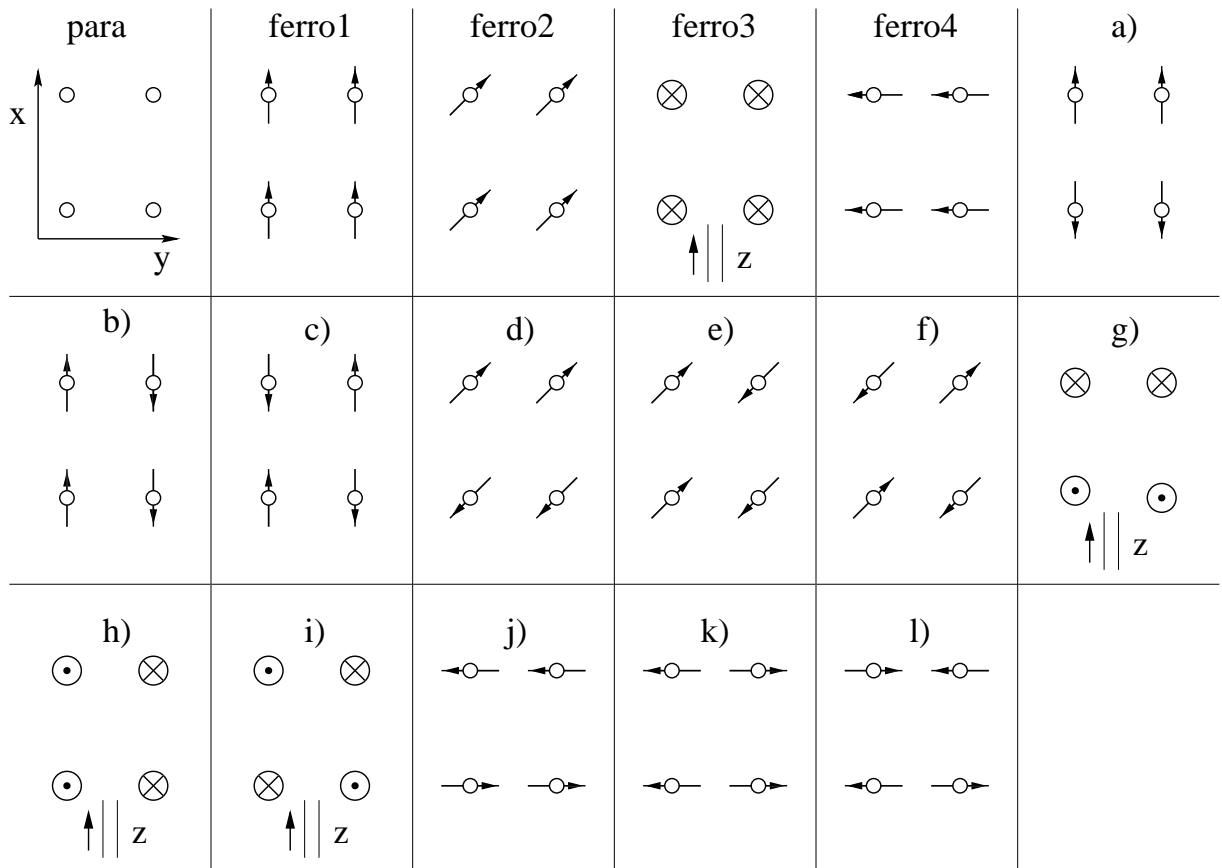, width=0.9\linewidth}
\caption{\label{f110}Spin configurations of an fcc (110) monolayer. Except for 
confs. ``ferro3", g), h), and i), the arrows always indicate in-plane directions 
of the spins. In confs. ``ferro3", g), h), and i) $\odot$ ($\otimes$) denote 
spins pointing along the positive (negative) z-direction, respectively.}
\end{figure}

\begin{figure}
\epsfig{file=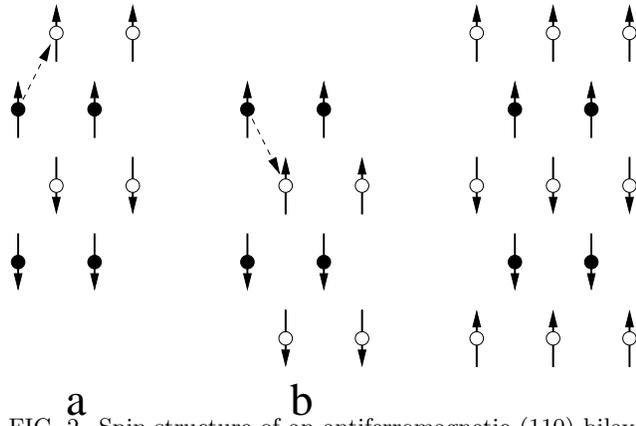, scale=0.4}
\caption{\label{cons110}Spin structure of an antiferromagnetic (110) bilayer 
constructed from a shift of the monolayer, where two different shiftings are 
applied. 
Filled (empty) circles represent the topmost (second) layer. The rightmost panel 
shows the conventional unit cell for the resulting bilayer structure. Here, 
conf. a) of the (110) surface serves as an example.}
\end{figure}

The configurations of the (110) monolayer structure are depicted in Fig. 
\ref{f110}, and the way the bilayer is constructed is depicted in Fig. 
\ref{cons110}. The tensor elements are presented in Table \ref{t110}. In 
general, we can observe five types of response. However, the possibility to 
distinguish AF configurations is not much improved compared to the (110) 
monolayer. Even the possibility to detect the magnetic phase of the surface is 
not evident. 

As for the (001) surface \cite{ourpaper}, there is no difference in SHG signal 
between the monolayer and bilayer for the paramagnetic and ferromagnetic phases. 
For most AF configurations, however (confs. a), b), c), e), f$_a$, f$_b$), g), 
h), j), k), and l)) such a difference is present due to a lower symmetry of the 
bilayer.

\subsection{(111) bilayer}

In order to be consistent with our previous work \cite{ourpaper} we keep the 
same configuration names as in this earlier paper. That is why, for example, 
conf. b) is not present here. The spin configurations of the (111) bilayer are 
constructed from the configurations of the (111) surface of our previous work in 
the way that the spin structure in the second atomic layer is the same as in the 
topmost layer, but shifted accordingly to form a hcp structure. Taking into 
account the spin structure of the second layer causes all the AF configurations 
to split, thus one obtains 10 AF configurations of the (111) bilayer. The 
configurations are labeled by small letters (indicating their ``parent'' 
configuration) with subscript ``a'' if the mentioned shifting is along the 
positive $x$ axis, and ``b'' if the shifting is along the negative S$_{xy}$ 
axis.

\begin{figure}
\epsfig{file=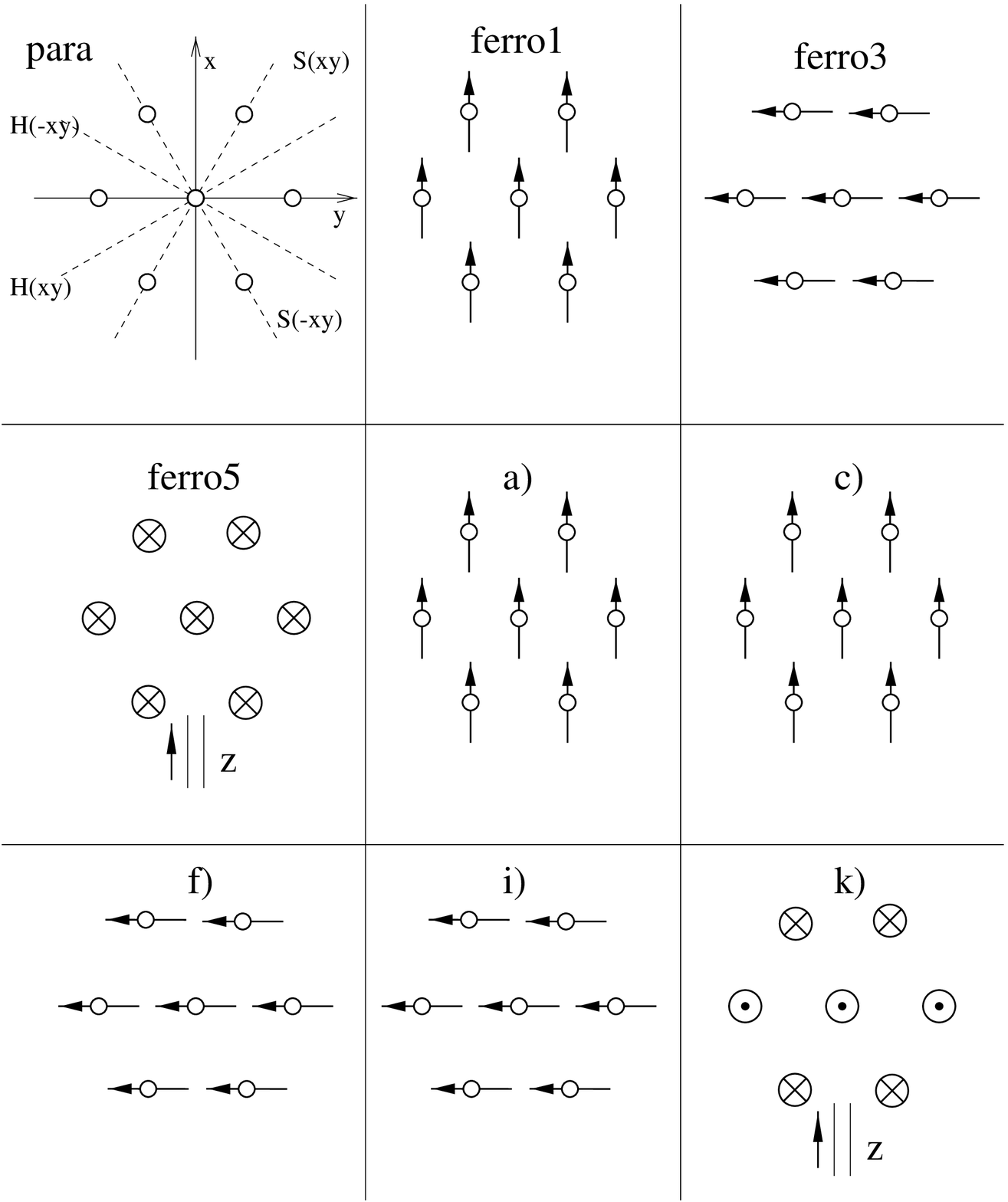, width=0.5\linewidth}
\caption{\label{f111}Spin configurations of an fcc (111) monolayer. Except for 
confs. ``ferro5", k), l), and m), the arrows always indicate in-plane directions 
of the spins. In confs. ``ferro5" and k) $\odot$ ($\otimes$) denote 
spins pointing along the positive (negative) z-direction, respectively.}
\end{figure}

\begin{figure}
\epsfig{file=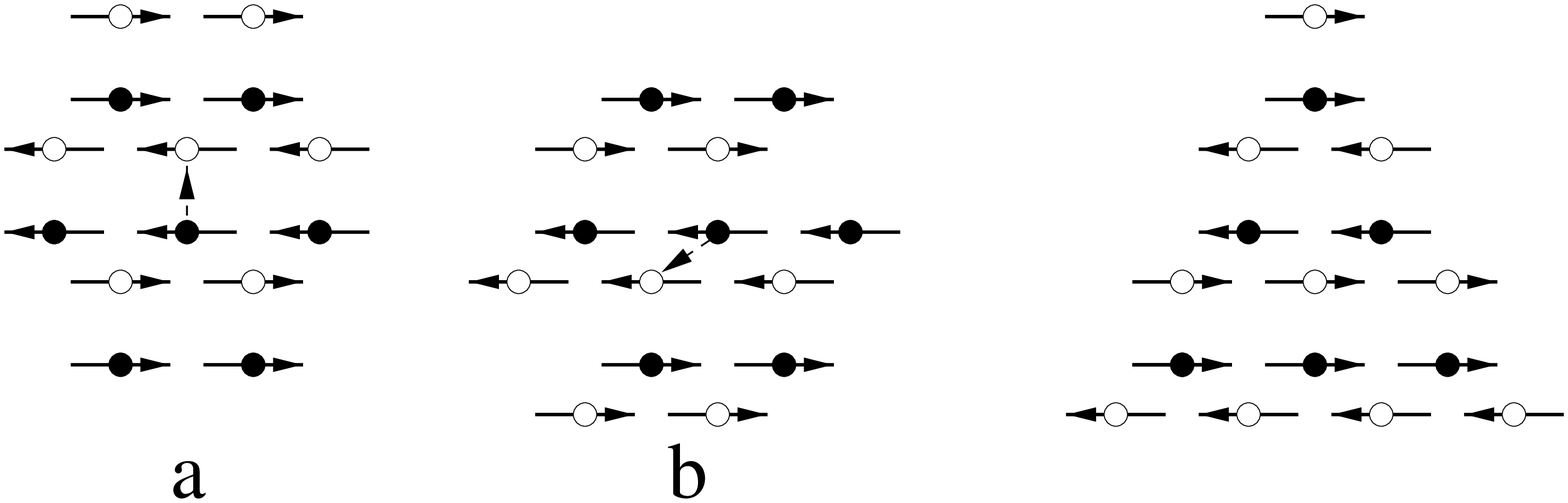, scale=0.4}
\caption{\label{cons111}Spin structure of an antiferromagnetic (111) bilayer 
constructed from a shift of the monolayer, where two different shiftings are 
applied. 
Filled (empty) circles represent the topmost (second) layer. Here, conf. f) of 
the (111) monolayer serves as an example. The rightmost panel displays the 
conventional unit cell for the resulting bilayer structure of conf. f$_a$).}
\end{figure}

The configurations of the (111) monolayer are depicted in Fig. \ref{f111} and 
the construction of the bilayer is depicted in Fig. \ref{cons111}. The 
corresponding tensor elements are displayed in Tab. \ref{t111}. The results are 
identical to those of our previous work \cite{ourpaper}, where the second layer 
of the (111) surface was treated as nonmagnetic. 
This means that the spin spin structure of the second layer does not play any 
role for SHG, however the presence of the atoms in the second layer does.

\section{Conclusion}
From our results follows that SHG can probe maximally two atomic layers of the 
surface of cubic two sublattice antiferromagnets, and only one of the 
paramagnetic or ferromagnetic surface. For the (111) surface, the {\em spin} 
structure of the second layer does not have any influence on SHG, i.e. it does 
not matter from the group-theoretical point of view if the investigated surface 
is a termination of a bulk antiferromagnet or a monolayer grown on a nonmagnetic 
substrate. However, these two situations can be very different from the 
band-theoretical point of view.

\end{multicols}

\newpage
\begin{table}
\caption[tab110A]{\label{t110}Nonvanishing elements of $\chi^{(2\omega)}_{el}$ 
for all spin configurations of the (110) surface of a fcc lattice.
We denote $\chi^{(2\omega)}_{ijk}$ by ijk. Odd elements are in bold if a parity 
operation exists. The configurations are depicted in Fig. \ref{f110}.} 
\begin{tabular}{cclcl}
conf.&point group&symmetry operations&domain operation&non-vanishing 
tensor elements\\ \hline
 & & & \\
para&mm2&$1,2_z,\overline{2}_x,\overline{2}_y$&-& xxz = xzx, yyz = yzy, zxx, 
zyy, zzz \\
ferro1&m&$1,\overline{2}_x$&$2_z,\overline{2}_y$& xzx = xxz, {\bf xxy} = {\bf 
xyx}, {\bf yxx}, {\bf yyy}, {\bf yzz}, \\
& & & & yyz = yzy, zxx, zyy, zzz, {\bf zyz} = {\bf zzy} \\
ferro2&1&1&$2_z$& All the elements are allowed: \\
&&&& {\bf xxx}, {\bf xyy}, {\bf xzz}, xyz = xzy, xzx = xxz, {\bf xxy} = {\bf 
xyx}, \\
&&&& {\bf yxx}, {\bf yyy}, {\bf yzz}, yyz = yzy, yzx = yxz, {\bf yxy} = {\bf 
yyx}, \\
&&&& zxx, zyy, zzz, {\bf zyz} = {\bf zzy}, {\bf zzx} = {\bf zxz}, zxy = zyx\\
&&&$\overline{2}_x$&{\bf xxx}, {\bf xyy}, {\bf xzz}, {\bf xyz} = {\bf xzy}, xzx 
= xxz, xxy = xyx, \\
&&&&yxx, yyy, yzz, yyz = yzy, {\bf yzx} = {\bf yxz}, {\bf yxy} = {\bf yyx}, \\
&&&&zxx, zyy, zzz, zyz = zzy, {\bf zzx} = {\bf zxz}, {\bf zxy} = {\bf zyx}\\
ferro3&2&$1,2_z$&$\overline{2}_x,\overline{2}_y$& {\bf xyz} = {\bf xzy}, xxz = 
xzx, yyz = yzy, {\bf yzx} = {\bf yxz}, \\
& & & &zxx, zyy, zzz, {\bf zxy} = {\bf zyx} \\
ferro4&m&$1,\overline{2}_y$&$2_z,\overline{2}_x$&{\bf xxx}, {\bf xyy}, {\bf 
xzz}, xxz = xzx, yyz = yzy, \\
&&&&{\bf yyx} = {\bf yxy}, zxx, zyy, zzz, {\bf zzx} = {\bf zxz}\\
AF: & & & & \\
a), g), j)&m&$1,\overline{2}_y$&$2_z,\overline{2}_x$&{\bf xxx}, {\bf xyy}, {\bf 
xzz}, xxz = xzx, yyz = yzy, \\
&&&&{\bf yyx} = {\bf yxy}, zxx, zyy, zzz, {\bf zzx} = {\bf zxz}\\
b), h), k)&m&$1,\overline{2}_x$&$2_z,\overline{2}_y$& xzx = xxz, {\bf xxy} = 
{\bf xyx}, {\bf yxx}, {\bf yyy}, {\bf yzz}, \\
& & & & yyz = yzy, zxx, zyy, zzz, {\bf zyz} = {\bf zzy} \\
c), d), l)&2&$1,2_z$&$\overline{2}_x,\overline{2}_y$& {\bf xyz} = {\bf xzy}, xxz 
= xzx, yyz = yzy, {\bf yzx} = {\bf yxz}, \\
& & & & zxx, zyy, zzz, {\bf zxy} = {\bf zyx} \\
e), f$_a$), f$_b$)&1&1&$2_z$& All the elements are allowed: \\
&&&& {\bf xxx}, {\bf xyy}, {\bf xzz}, xyz = xzy, xzx = xxz, {\bf xxy} = {\bf 
xyx}, \\
&&&& {\bf yxx}, {\bf yyy}, {\bf yzz}, yyz = yzy, yzx = yxz, {\bf yxy} = {\bf 
yyx}, \\
&&&& zxx, zyy, zzz, {\bf zyz} = {\bf zzy}, {\bf zzx} = {\bf zxz}, zxy = zyx\\
&&&$\overline{2}_x$&{\bf xxx}, {\bf xyy}, {\bf xzz}, {\bf xyz} = {\bf xzy}, xzx 
= xxz, xxy = xyx, \\
&&&&yxx, yyy, yzz, yyz = yzy, {\bf yzx} = {\bf yxz}, {\bf yxy} = {\bf yyx}, \\
&&&&zxx, zyy, zzz, zyz = zzy, {\bf zzx} = {\bf zxz}, {\bf zxy} = {\bf zyx}\\
i$_a$), i$_b$)&mm2&$1,2_z,\overline{2}_x,\overline{2}_y$&one domain& xxz = 
xzx, yyz = yzy, zxx, zyy, zzz \\
\end{tabular}
\end{table}   

\begin{table}
\caption[tab1112A]{\label{t111}Nonvanishing elements of $\chi^{(2\omega)}_{el}$ 
for all spin configurations of the (111) surface of a fcc lattice.
More monolayers are taken into account. We denote $\chi^{(2\omega)}_{ijk}$ by 
ijk. The configurations are depicted in Fig. \ref{f111}.} 
\begin{tabular}{cclcl}
conf.&point grp.&symmetry ops.&domain op.&non-vanishing 
tensor elements\\ \hline
 & & & & \\
para&3m&$1,\pm3_z,\overline{2}_y,\overline{2}_{S(xy)},\overline{2}_{S(-xy)}$
&-&zxx = zyy, xxz = xzx = yyz = yzy, zzz, \\
& & & & xxx = -xyy = -yxy = -yyx \\
ferro1&1&1&$\overline{2}_y$& All the elements are allowed: \\
&&&& xxx, xyy, xzz, {\bf xyz} = {\bf xzy}, xzx = xxz, {\bf xxy} = {\bf xyx}, \\
&&&& {\bf yxx}, {\bf yyy}, {\bf yzz}, yyz = yzy, {\bf yzx} = {\bf yxz}, yxy =  
yyx, \\
&&&& zxx, zyy, zzz, {\bf zyz} = {\bf zzy}, zzx = zxz, {\bf zxy} = {\bf zyx}\\
ferro3&m&$1,\overline{2}_y$&-& xxx, xyy, xzz, xxz = xzx, yyz = yzy, \\
& & & & yyx = yxy, zxx, zyy, zzz, zzx = zxz \\
ferro5&3&$1,\pm3_z$&$\overline{2}_y$& xxx = -xyy = -yxy = -yyx, {\bf xyz} = {\bf 
xzy} = -{\bf yxz} = -{\bf yzx},\\
&&&& xzx = xxz = yyz = yzy, {\bf xxy} = {\bf xyx} = {\bf yxx} = -{\bf yyy},\\
&&&& zxx = zyy, zzz\\
AF:& & & &\\
$\!\!\!$a$_a$), a$_b$), i$_a$),&m&$1,\overline{2}_y$&-& xxx, xyy, xzz, xxz = 
xzx, yyz = yzy, \\
$\;\;\;$ i$_b$), k$_a$), k$_b$)& & & & yyx = yxy, zxx, zyy, zzz, zzx = zxz \\
$\!\!\!$c$_a$), c$_b$),&1&$1$&$\overline{2}_y$& All the elements are allowed: \\
$\;\;\;$ f$_a$), f$_b$)&&&& xxx, xyy, xzz, {\bf xyz} = {\bf xzy}, xzx = xxz, 
{\bf xxy} = {\bf xyx},\\
&&&& {\bf yxx}, {\bf yyy}, {\bf yzz}, yyz = yzy, {\bf yzx} = {\bf yxz}, yxy = 
yyx,\\
&&&& zxx, zyy, zzz, {\bf zyz} = {\bf zzy}, zzx = zxz, {\bf zxy} = {\bf zyx}\\
\end{tabular}
\end{table}   

\begin{thebibliography} {99}
\bibitem{par111} The notion of the parity operation requires that the original 
situation is restored after applying the operation at most twice. Obviously, 
$2_z$ is such an operation, and for certain configurations $4_z$ satisfies this 
criterion as well (where $2_z$ is just a symmetry operation). However, neither 
$3_z$ nor $6_z$ have this property - they must be applied at least 3 times to 
restore the original situation. Concerning the tensor elements, for $2_z$ and 
mirror-operations a given tensor element may change its sign or not. Upon the 
application of the $4_z$ operation it may become another tensor element (i.e. 
its x indices become y and vice versa) while its sign can change or remain 
unchanged. Upon the $3_z$ operation, however, a single tensor element gets 
``split'' into several tensor elements, and the notion of the conservation (or 
not) of the sign of the tensor element looses its sense. The tensor elements are 
described in the cartesian coordinate system, where quarters are the elementary 
entities. Rotation by an angle other than $90^\circ$ and its multiples cannot be 
described as interchanging the axes and a possible modification of their signs. 
However, it is possible to treat the parity of tensor elements in the $3_z$ and 
$6_z$ operation in a more complicated way.
\end{thebibliography}
\end{document}